\documentclass[11pt]{article}
\usepackage{amsmath, amsthm, amssymb}
\usepackage{graphicx}
\usepackage{natbib}
\newcommand{\yay}{\textsf{True}}
\newcommand{\nay}{\textsf{False}}

\newcommand{\str}[1]{\texttt{"#1"}}

\newtheorem{definition}{Definition}

\title{Data Validation}
\author{Mark P.J. van der Loo\footnote{Corresponding author: mplo@cbs.nl} and Edwin de Jonge\\
Statistics Netherlands}

\date{This is the submitted version of the following book chapter: ``Van der
Loo, M.P.J. and De Jonge, E. (2020) \emph{Data Validation}. In Wiley StatsRef:
Statistics Reference Online'' which has been published in final form in\\
\texttt{https://onlinelibrary.wiley.com/doi/10.1002/9781118445112.stat08255}.
}

\begin{document}
\maketitle
\begin{abstract}
Data validation is the activity where one decides whether or not a particular
data set is fit for a given purpose. Formalizing the requirements that drive
this decision process allows for unambiguous communication of the requirements,
automation of the decision process, and opens up ways to maintain and
investigate the decision process itself. The purpose of this article is to
formalize the definition of data validation and to demonstrate some of the
properties that can be derived from this definition. In particular, it is shown
how a formal view of the concept permits a classification of data quality
requirements, allowing them to be ordered in increasing levels of complexity.
Some subtleties arising from combining possibly many such requirements are
pointed out as well.
\end{abstract}

Informally, data validation is the activity where one decides whether or not a
particular data set is fit for a given purpose. The decision is based on
testing observed data against prior expectations that a plausible dataset is
assumed to satisfy. Examples of prior expectations range widely. They include
natural limits on variables (weight cannot be negative), restrictions on
combinations of multiple variables (a man cannot be pregnant), combinations of
multiple entities (a mother cannot be younger than her child) and combinations
of multiple data sources (import value of country A from country B must equal
the export value of country B to country A). Besides the strict logical
constraints mentioned in the examples, there are often `softer' constraints
based on human experience. For example, one may not expect a certain economic
sector to grow more than 5\% in a quarter. Here, the 5\% limit does not
represent a physical impossibility but rather a limit based on past experience.
Since one must decide in the end whether a data set is usable for its intended
purpose, we treat such assessments on equal footing.

The purpose of this paper is to formalize the definition of data validation and
to demonstrate some of the properties that can be derived from this definition.
In particular, it is shown how a formal view of the concept permits a
classification of data validation rules (assertions), allowing them to be
ordered in increasing levels of `complexity'. Here, the term `complexity'
refers to the amount of different types of information necessary to evaluate a
validation rule. A formal definition also permits development of tools for
automated validation and automated reasoning about data validation
\citep{zio2015methodology, loo2018statistical, loo2019infrastructure}. Finally,
some subtleties arising from combining validation rules are pointed out.

\section{Formal definition of data validation}
Intuitively, a validation activity classifies a dataset as acceptable or not
acceptable.  A straightforward formalization is to define it as a function from
the collection of data sets that could have been observed, to
$\{\yay{},\nay{}\}$. One only needs to be careful in defining the `collection
of data sets', to avoid a `set of all sets' which recursively holds itself. To
avoid such paradoxes, a data set is defined as a set of key-value pairs, where
the keys come from a finite set, and the values from some domain.
\begin{definition}
\label{def:point}
A \emph{data point} is a pair $(k, x)\in K\times D$, where $k$ is a \emph{key},
selected from a finite set $K$ and $x$ is a \emph{value} from a \emph{domain}
$D$.
\end{definition}
In applications the identifier $k$ makes the value interpretable as the
property of a real-world entity or event. The domain $D$ is the set of all
possible values that $x$ can take, and it therefore depends on the
circumstances in which the data is obtained. 

As an example, consider an administrative system holding age and job status of
persons. It is assumed that `job' takes values in $\{\str{employed},
\str{unemployed}\}$ and that `age' is an integer.  However, if the data entry
system performs no validation on entered data, numbers may end up in the job
field, and job values may end up in the age field.  In an example where the
database contains data on two persons identified as $1$ and $2$, this gives
\begin{align}
\begin{split}
K &= \{1,2\}\times\{\str{age}, \str{job}\}\\
D &= \mathbb{N}\cup \{\str{employed},\str{unemployed}\}.
\end{split}
\label{eq:example}
\end{align}
This definition allows for the occurrence of missing values by defining a
special value for them, say `\textsf{NA}' (not available) and adding it to the
domain.

Once $K$ and $D$ are fixed it is possible to define the set of all
observable datasets.
\begin{definition}
A \emph{dataset} is a subset of $K\times D$ where every key in $K$
occurs exactly once.
\end{definition}
Another way to interpret this is to say that a data set is a total function
$K\to D$. The set of all observable data sets is denoted $D^K$. In the example,
one possible dataset is
\begin{align*}
  & \{((1,\str{age}), 25), ((1,\str{job}), \str{unemployed}),\\
  &\phantom{\{} ((2,\str{age}), \str{employed}), ((2,\str{job}), 42)\}
\end{align*}
Observe that the key consists of a person identifier and a variable identifier.
Since type checking is a common part of data validation these definitions
deliberately leave open the possibility that variables assume a value of the
wrong type. 

Data validation can now be formally defined as follows.
\begin{definition}
\label{def:valifun}
A \emph{data validation function} is a surjective function 
$$
v: D^K\twoheadrightarrow \{\nay{},\yay{}\}.
$$
\end{definition}
A data set $d\in D^K$ for which $v(d)=\yay{}$ is said to \emph{satisfy} $v$.  A
data set for which $v(d)=\nay{}$ is said to \emph{fail} $v$.

Recall that surjective means that there is at least one $d\in D^K$ for which
$v(d)=\nay{}$ and at least one $d\in D^K$ for which $v(d)=\yay{}$.  A
validation function has to be surjective to have any meaning as a data
validation activity. Suppose  $v':D^K\to \{\nay{},\yay{}\}$ non-surjective
function. If there is no data set $d$ for which $v'(d)=\nay{}$, then $v'$ is
always true and it does not validate anything. If, on the other hand, there is
no $d$ for which $v'(d)=\yay{}$ then no data set can satisfy $v'$. In short, a
function that is not surjective on $\{\nay{}, \yay{}\}$ does not separate valid
from invalid data.

A data validation function is reminiscent of a predicate as defined in
first-order logic. Informally, a predicate is a statement about variables that
can be \yay{} or \nay{}. The variables can take values in a predefined set
(referred to as the \emph{domain} of the predicate). Since validation functions
map elements of $D^K$ to $\{\nay{}, \yay{}\}$, it is tempting to equate a
validation function as a predicate over $D^K$. However, the elements of $D^K$
are instances of possible data sets, and validation is based on statements
involving variables of a single observed data set. It is therefore more
convenient to adopt the following definition.
\begin{definition}
A \emph{validation rule} is a predicate over an instance $d\in D^K$ that is
neither a tautology nor a contradiction.
\label{def:validationrule}
\end{definition}
The elements of $D^K$ are sufficiently similar so that any validation rule over
a particular data set $d\in D^K$ is also a validation rule over another data
set in $d'\in D^K$, where the truth value of a validation rule depends on the
chosen data set. This also allows us to interpret a tautology as a predicate
that is \yay{} for every element of $D^K$ and a contradiction as a predicate
that is \nay{} for every element of $D^K$.  

The equivalence between an assertion about a data set and a function
classifying possible data sets as valid or invalid instances is a rather
obvious conclusion. The actual value of the above exercise is the strict
definition of a data point as a key-value pair. As will be shown below,
inclusion of the key permits a useful classification of data validation rules.

To demonstrate that many types of validation rules can be expressed in this
framework a few examples based on the example of Equation~\eqref{eq:example}
will be considered. The following rule states that all ages must be integer.
$$
\forall ((u,\str{age}),x)\in d: x \in \mathbb{N}.
$$
Here, $((u,\str{age}),x)$ runs over the person-value pairs where the value is
supposed to represent an age. Similarly, it is possible express a nonnegativity
check on the age variable.
$$
\forall ((u,\str{age}),x)\in d: x \geq 0.
$$
In these examples a data set fails a validation rule when not all elements
satisfy a predicate. Such rules are not very informative when it comes to
pinpointing what elements of the data set cause the violation. It is customary
to perform data validation on a finer level, for example by checking 
nonnegativity element by element. Based on the definitions introduced here
this is done by fixing the key completely.
\begin{align*}
\forall ((1,\str{age}),x)\in d: x\geq 0\\
\forall ((2,\str{age}),x)\in d: x\geq 0.
\end{align*}
Now consider the cross-variable check that states that employed persons must be
$15$ or older.
\begin{align*}
\forall ((u,\str{age}),x), (u,\str{job}),y)\in d: y=\str{employed}\Rightarrow x\geq 15,
\end{align*}
where $\Rightarrow$ denotes logical implication (if $y=\str{employed}$ then
$x\geq 15$).  Finally, consider the cross-person check that states that we
expect the average age in the data set to be larger then or equal to 5.
$$
\frac{
\sum_{(u,\str{age},x)\in d} x}{
\sum_{(u,X,x)\in d} \delta(X,\str{age})
} \geq 5,
$$
where $\delta(X,Y)=1$ when $X=Y$ and otherwise 0. 

In practical applications validation rules are often expressed more directly in
terms of variable names, such as $age \geq 0$. Such expressions are
specializations where one silently assumes extra properties such as that the
rule will be evaluated for the entry for age in every record. 

\paragraph{Remark 1.} In Definition~\ref{def:valifun}, (and
also~\ref{def:validationrule}) a validation function is defined as a surjection
$D^K\twoheadrightarrow\{\nay{},\yay{}\}$. In practical applications it is often
useful to also allow the value \texttt{NA} (not available) for cases where one
or more of the data points necessary for evaluating the validation rule are
missing. In that case the domain $D$ in the Equation~\eqref{eq:example} must be
extended to $D\cup \{\texttt{NA}\}$. See~\citet{loo2019infrastructure} for an
implementation.

\paragraph{Remark 2.} The current formalization of data validation excludes
checking for completeness of the key set: it is assumed by definition that each
data set in $D^K$ is a complete set of key-value pairs where the keys cover all
of $K$. The above definitions therefore create a clean distinction between what
is metadata (the keys and their interpretation) and data (key-value pairs). It
is possible to check for uniqueness of variables.

\paragraph{Remark 3.} The formal definition of data validation rules also
allows a formalization of data validation software tools or domain specific
languages, such as in \cite{loo2019infrastructure}.

\paragraph{Remark 4.} In official (government) statistics, validation rules are
referred to as \emph{edit rules} rather than data validation rules. See
\citet{waal2011handbook} and references therein.

\section{Classification of validation rules}
Validation rules defined by domain experts may be easy for humans to interpret,
but can be complex to implement. Take as an example the assertion `the average
price of a certain type of goods this year, should not differ more then ten
percent from last year's average'. To evaluate this assertion one needs to
combine prices of multiple goods collected over two time periods. A practical
question is therefore if the `complexity of evaluation', in terms of the amount
of information necessary, can somehow be measured. In this section a
classification of data validation rules is derived that naturally leads to an
ordering of validation rules in terms of the variety of information that is
necessary for their evaluation.

One way to measure the amount of information is to look at the predicate
defining a data validation rule and to determine how many different $(k,x)$
pairs are needed to evaluate it. This is not very useful for comparing
complexity across different data sets that are not from the same $D^K$ since
the numbers will depend on the size of the key set $K$. One measure that does
generalize to different $D^K$ is the measure `does a rule need one, or several
$(k,x)$ to be evaluated?'.

This measure is not very precise, but it can be refined when a key consists of
multiple meaningful parts such as in the running example where the key consists
of the id of a person and the name of a variable.  One can then classify a rule
according to two questions: `are multiple person id's necessary for
evaluation?', and: `are multiple variables necessary for evaluation?'. This
gives a four-way classification: one where both questions are answered with
`no', two where one of the questions is answered with `yes' and one where both
questions are answered with `yes'.  Although this refinement improves the
accuracy of the classification, it only allows for comparing validation rules
over data sets with the exact same key structure. It would therefore be useful
to have a generic key structure that can be reused in many situations. One such
structure can be found by considering in great generality how a data point is
obtained in practice.

\begin{figure}
\centering
\includegraphics[width=0.5\textwidth]{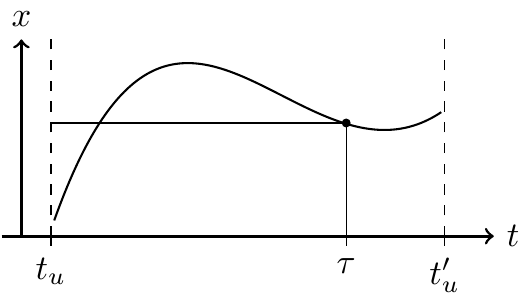}
\caption{
A unit $u$ of type $U$ exists from $t_u$ to $t'_u$. From $t_u$
onward it has an attribute $X$ with its value $x$ possibly changing over time.
At some time $\tau$ this value is observed. The observed value is thus fully
characterized by the quartet ($U$, $\tau$,$u$, $X$).
}
\label{fig:measurement}
\end{figure}
A data point usually represents the value of an attribute of an object or event
in the real world: a person, a web site, an e-mail, a radio broadcast, a
country, a sensor, a company, or anything else that has observable attributes.
In what follows, an object or event is referred to as a `unit' for brevity.  A
data point is created by observing an attribute $X$ from a unit $u$ of a
certain type $U$ at time $\tau$, as in Figure~\ref{fig:measurement}. Using the
same reasoning as above,  this yields a $2^4=16$-way classification of
validation rules: for each element $U$, $\tau$, $u$, $X$ a validation rule
requires a single or multiple instances to be evaluated. However, there are some
restrictions.  Any unit $u$ can only be of one type. So evaluating a validation
rule will never require multiple types and a single unit. Second, the type of a
unit fixes its properties. So a validation rule will never need a single
variable for multiple types. With these restrictions considered, the number
of possible classes of validation rules reduces from sixteen to ten.

To distinguish the classes the following notation is introduced. For each
element $U$, $\tau$, $u$, $X$ assign an $s$ when a validation rule pertains to
a single value of that element and assign an $m$ when a validation rule
pertains to multiple values of that element.  For example, a validation rule of
class $sssm$ needs a single type, a single measurement, a single object and
multiple variables to be evaluated. 

The ten possible classes can themselves be grouped into \emph{validation levels},
according to whether a class is labeled with no, one, two, three, or four
$m$'s.  A higher validation level indicates that we need a larger variety of
information in order to evaluate the rule. The classification, and their
grouping into validation levels is summarized in Table~\ref{tab:validationlevels}.
\begin{table}[t]
\caption{The ten possible classes of validation rules, grouped into `validation levels'.
A higher level indicates that a wider variety of information is necessary to
evaluate a validation rule.}
\label{tab:validationlevels}
\centering
\begin{tabular}{@{}ccccc@{}}
\multicolumn{5}{c}{Validation level}\\
\hline
0 & 1 & 2 & 3 & 4\\
\hline
ssss & sssm & ssmm & smmm & mmmm \\
     & ssms & smsm & msmm &      \\
     & smss & smms &      &      \\
\hline
\end{tabular}
\end{table}

Going from level 0 to 4 corresponds to a workflow that is common in
practice. One starts with simple tests such as range checks.  These are of
level zero since a range check can be performed on a single data point. That
is, one only needs a value that corresponds to a single type, measurement,
unit, and variable.  Next, more involved checks are performed, for instance,
involving multiple variables ($sssm$, e.g. the ratio between two properties of
the same unit must be in a given range), multiple units ($ssms$, e.g.  the mean of
a variable over multiple units must be within a specified range), or multiple
measurements ($smss$, e.g.  the current value of the property of a unit cannot
differ too much from a past value of the same property of the same unit).
Going up in level, even more complex rules are found until rule evaluation
involves multiple variables of multiple units of multiple types measured at
multiple instances ($mmmm$). 

This classification also has an immediate interpretation for data stored in a
data base that is normalized in the sense of \citet{codd1970relational}. In
such a data base, records represent units, columns represent variables, and tables
represent types. The `time of measurement' is represented as a variable as
well. The classification indicates whether a rule implementation needs to
access multiple records, columns or tables.

\section{Properties of validation rule sets}
Definition~\ref{def:validationrule} implies that a validation rule is a
predicate over a data set that is not a tautology nor a contradiction. This
means that combining two validation rules with $\land$ or $\lor$ does not
automatically yield a new validation rule.  Consider for example the rules
$x\geq 0$ and $x \leq 1$ (using shorter notation for brevity). The rule $x \geq
0 \lor x\leq 1$ is a tautology.  The rule $x \geq 0 \land x\leq -1$ is a
contradiction. In fact, the only operation that is guaranteed to transform a
validation rule into another validation rule is negation.

The fact that validation rules are not closed under conjugation ($\land$) or
disjunction ($\lor$) has practical consequences. After all, defining a set of
validation rules amounts to conjugating them together into a single rule since
a data set is valid only when all validation rules are satisfied. A set of
rules may be such that their conjugation is a contradiction.  Such a rule set
is called \emph{infeasible}. More subtle problems involve unintended
consequences, including \emph{partial infeasibility} \citep{bruni2012formal},
and introduction on fixed values or range restrictions. Other problems involve
the introduction of several types of redundancies \citep{dillig2010small,
paulraj2010comparative}, which make rule sets both harder to maintain and
hamper solving problems such as error localization \citep{bruni2005error,
jonge2019errorlocalization}. In the following, some examples of unintended
effects and redundancies are discussed.  The examples shown here are selected
from a more extensive discussion in \citet[Chapter 8]{loo2018statistical}.  For
simplicity of presentation the rules are expressed as simple clauses,
neglecting the key-value pair representation.

Partial inconsistency is (often) an unintended consequence implied by a pair
of rules. For example the rule set
\begin{align*}
gender &= \str{male} \Rightarrow income > 2000\\
gender &= \str{male} \Rightarrow income < 1000,
\end{align*}
is feasible, but it can only be satisfied when $gender\not=\str{male}$.  Thus,
the combination of rules (unintentionally) excludes an otherwise valid gender
category. 

A simple redundancy is introduced when one rule defines a subset of valid
values with respect to another rule. For example if $x \geq 0$ and $x\geq 1$,
then $x\geq 0$ is redundant with respect to $x\geq 1$. More complex cases arise
in sets with multiple rules. A more subtle redundancy, called `nonrelaxing clause'
occurs in the following situation.
\begin{align*}
x \geq 0 \Rightarrow y \geq 0\\
x \geq 0.
\end{align*}
Here, the second rule implies that the condition in the first rule must always
be true. Hence the rule set can be simplified to 
\begin{align*}
y \geq 0\\
x \geq 0.
\end{align*}
Another subtle redundancy, called a `nonconstraining clause' occurs in
the following situation.
\begin{align*}
x > 0 &\Rightarrow y > 0\\
x < 1 &\Rightarrow y > 1
\end{align*}
Now, letting $x$ vary from $-\infty$ to $\infty$, the rule set first implies
that $y>1$. As $x$ becomes positive, the rule set implies that $y>0$ until $x$
reaches $\infty$. In other words, the rule set implies that $y$ must be
positive regardless of $x$ and can therefore be replaced with
\begin{align*}
y > 0\\
x < 1 &\Rightarrow y > 1
\end{align*}
Methods for algorithmically removing such issues and simplifying rule sets have
recently been discussed by \cite{daalmans2018constraint}, and have been
implemented by \citet{jonge2019validatetools}. In short, the methods are based
on formulating Mixed Integer Programming (MIP) problems and detecting their
(non)-convergence after certain manipulations of the rule sets. General theory
and methods for rule manipulation have also been discussed by
\citet{chandru1999optimization}  and \citet{hooker2000logic} in the context of
optimization.

\section{Conclusion}
Data validation can be formalized equivalently in terms of certain predicates
over a data set or as a surjective Boolean function over a well-defined set of
observable data sets. It is possible to define a very general classification of
validation rules, based on a generic decomposition of the metadata. Combining
validation rules into a set can lead to subtle and unintended consequences that
can be solved in some cases with algorithmic methods.

\section*{Related articles}

\begin{itemize}
\item stat04233 
\item stat06088 
\item stat06776 
\item stat04018 
\end{itemize}

\bibliography{statref}

\end{document}